\RequirePackage{fix-cm}
\documentclass[dvipdfmx]{svjour3}                     
\smartqed  
\usepackage[dvipdfmx]{graphicx}
\usepackage{amsmath,amssymb}
\usepackage{amsfonts}
\usepackage{physics}
\usepackage{hyperref}
\usepackage{bm}
\usepackage[title]{appendix}
\usepackage[subrefformat=parens]{subcaption}
\newcommand{\spstate}[1]{\mathsf{#1}}
\begin{document}
\title{{\bf Ferromagnetism in $d$-dimensional SU($n$) Hubbard models with nearly flat bands}
}


\author{Kensuke Tamura \and
Hosho Katsura 
}


\institute{Kensuke Tamura \at
              Department of Physics, Graduate School of Science. The University of Tokyo, 7-3-1 Hongo, Tokyo 113-0033, Japan \\
              \email{tamura-kensuke265@g.ecc.u-tokyo.ac.jp}  
           \and
           H. Katsura \at
             Department of Physics, Graduate School of Science. The University of Tokyo, 7-3-1 Hongo, Tokyo 113-0033, Japan\\
             Institute for Physics of Intelligence, The University of Tokyo, 7-3-1, Hongo, Tokyo 113-0033, Japan\\
           Trans-scale Quantum Science Institute, The University of Tokyo, 7-3-1, Hongo, Tokyo 113-0033, Japan
}

\date{Received: date / Accepted: date}

\maketitle

\begin{abstract}
We present rigorous results for the SU($n$) Fermi-Hubbard models with finite-range hopping in $d$ ($\ge 2$) dimensions.
The models are defined on a class of decorated lattices. 
We first study the models with flat bands at the bottom of the single-particle spectrum and prove that the ground states exhibit SU($n$) ferromagnetism when the number of particles is equal to the number of unit cells. 
We then perturb the models by adding particular hopping terms and make the bottom bands dispersive.
Under the same filling condition, it is proved that the ground states remain SU($n$) ferromagnetic when the bottom bands are sufficiently flat and the Coulomb repulsion is sufficiently large.
\end{abstract}
\section{Introduction} \label{sec:1}
Recent advances in experimental techniques have made it possible to simulate various quantum systems by using ultracold atoms in optical lattices~\cite{bloch2008many,bloch2012quantum,lewenstein2012ultracold,ueda2010fundamentals}.
Thanks to the controllability of lattice potentials and interaction strengths with high accuracy, ultracold atomic systems are expected to be a versatile tool for exploring many-body physics in strongly correlated systems.
Of particular interest are multicomponent fermionic systems with alkaline earth-like atoms in cold-atom setups.
Experimental realizations of such systems with SU($n$) symmetric interactions have been reported in~\cite{taie20126,hofrichter2016direct,ozawa2018antiferromagnetic}.
They are expected to be well described by the SU($n$) Fermi-Hubbard model, which is a generalization of the standard Hubbard model with SU(2) symmetry~\cite{kanamori1963electron,gutzwiller1963effect,hubbard1963electron}, a minimal model for describing the properties of correlated electrons in solids.
In conventional condensed matter physics, the SU($n$) Hubbard model has been studied in the context of the large-$n$ approach~\cite{affleck1988large,marston1989large}.
In this approach, the main focus was on the large-$n$ limit, and the physical properties of the models at finite $n$ have been less investigated.
However, the recent experimental realizations of the SU($n$) Hubbard models with ultracold fermionic atoms have generated renewed theoretical interest in the study of the model at finite $n$ ($> 2$).
A number of studies revealed that the models can exhibit exotic phases that do not appear in the SU(2) counterpart~\cite{honerkamp2004ultracold,honerkamp2004bcs,rapp2007color,rapp2008trionic,cazalilla2009ultracold,cazalilla2014ultracold,capponi2016phases,chung20193}.
\par
Despite its apparent simplicity, mathematically rigorous treatment of the SU($n$) Hubbard models is, in general, a formidable task due to the intricate competition between the kinetic and the on-site Coulomb terms. 
Also, since the internal degrees of freedom will be larger compared to the SU(2) case, exact results for the SU($n$) Hubbard model are limited to fewer examples than in the SU(2) case.
The Nagaoka ferromagnetism was the first rigorous result for the SU(2) Hubbard model~\cite{nagaoka1966ferromagnetism,tasaki1989extension,thouless1965exchange}.
Given infinitely large Coulomb repulsions and exactly one hole, it was proved that the ground state of the Hubbard model defined on a lattice satisfying a certain connectivity condition is ferromagnetic and unique.
The Nagaoka ferromagnetism in the SU($n$) Hubbard model has also been established in~\cite{katsura2013nagaoka,bobrow2018exact}.
As for the multiorbital Hubbard models, theorems about ferromagnetism have been proved in~\cite{li2014exact,li2015exact}, and its extension to the SU($n$) case was also discussed~\cite{li2014exact}.
The above results are for singular cases in the sense that the Coulomb interaction is infinitely large.
The SU(2) Hubbard models with flat bands provide us with another rigorous example of ferromagnetism~\cite{mielke1991ferromagnetic,tasaki1992ferromagnetism,mielke1993ferromagnetism,tasaki1998nagaoka,tasaki2003ferromagnetism,mielketasaki1993ferromagnetism,tanaka2020extension,katsura2010ferromagnetism}.
Here, by a flat band, we mean a structure of a single-particle energy spectrum with a macroscopic degeneracy.
There are systematic methods for constructing tight-binding models with flat bands, such as the line graph~\cite{mielke1991ferromagnetic} and the cell construction~\cite{tasaki1992ferromagnetism}. 
In the SU(2) case, one can prove that the ground states are ferromagnetic and unique apart from the trivial spin degeneracy under the following conditions: (i) the number of particles is equal to the multiplicity of the single-particle ground states, and (ii) the basis for the space spanned by the single-particle ground states is connected.
Furthermore, its extension to the SU($n$) case was recently been discussed in~\cite{liu2019flat,tamura2019ferromagnetism}.
The flat-band ferromagnetism can also be thought of as a result for a singular situation because the density of states at the Fermi level diverges.
In the SU(2) case, the stability of flat-band ferromagnetism under perturbations which makes the lowest band dispersive has been discussed~\cite{tasaki1994stability,tasaki1996stability,tanaka2003stability,tanaka2018ferromagnetism}.
It was rigorously proved for a class of perturbed models in any dimension that the ground states remain ferromagnetic when the Coulomb repulsion and the band gap are sufficiently large~\cite{tasaki1995ferromagnetism,tasaki2020physics}.
By contrast, the stability of the flat-band ferromagnetism in the SU($n$) Hubbard model has been proved only in the one-dimensional case~\cite{tamura2019ferromagnetism}.
Thus, the generalizations in higher dimensions remain to be established.
In addition, since the Hohenberg-Mermin-Wagner theorem~\cite{mermin1966absence,hohenberg1967existence} forbids spontaneous symmetry breaking in one- and two-dimensional models with continuous symmetries at finite temperature, it is necessary to study models in dimensions higher than two in order to investigate ferromagnetism stable at finite temperature.
\par
In this paper, we study a class of SU($n$) Hubbard models on  $d$ ($\ge 2$)-dimensional decorated lattices and establish rigorous results. 
We first consider the models with flat bands at the bottom of the single-particle spectrum and prove that they exhibit SU($n$) ferromagnetism in their ground states, provided that the on-site Coulomb interaction is repulsive and the total fermion number is the same as the number of unit cells.
We then discuss SU($n$) ferromagnetism in perturbed models obtained by adding extra hopping terms that make the flat bands dispersive.
We prove that the particular perturbation leaves the ground states SU($n$) ferromagnetic when the band width of the bottom band is sufficiently narrow and the Coulomb repulsion is sufficiently large.
To establish the theorem for general $n$ and dimensions $d$, it is necessary to treat two cases, $n \le d$ and $n > d$, separately.
This is in marked contrast to the SU(2) case where the number of internal degrees of freedom cannot be greater than $d$ $(\ge 2)$. 
In addition, it should also be mentioned that our proof of Theorem 2 considerably simplifies the  previous proof for the nearly-flat-band ferromagnetism in the SU(2) case.
\par
The present paper is organized as follows.
In Sect.~\ref{sec:2}, we shall describe our model on a $d$-dimensional decorated hypercubic lattice and state our results about SU($n$) ferromagnetism.
In Sect.~\ref{sec:3}, we prove the first result for the SU($n$) flat-band ferromagnetism.
In Sect.~\ref{sec:4}, we prove our main result for the SU($n$) ferromagnetism in the model with a nearly flat band.
In Appendix~\ref{app:1}, we show the linear independence of many-body states defined by localized states, and in Appendix~\ref{app:2}, we give explicit expressions for the local ground states of the effective Hamiltonian in each particle sector. 
\section{Model and Main Results} \label{sec:2}
The models we consider are straightforward extensions of those in~\cite{tasaki2003ferromagnetism} to  SU($n$) case, and we follow the notation in~\cite{tasaki2003ferromagnetism}.
Readers are also referred to~\cite{tasaki1995ferromagnetism} for the original result about ferromagnetism of the SU(2) Hubbard model with nearly flat bands.
\subsection{Lattice} \label{subsec:lattice}
\begin{figure*}
	\centering
	\includegraphics[width=0.75\textwidth]{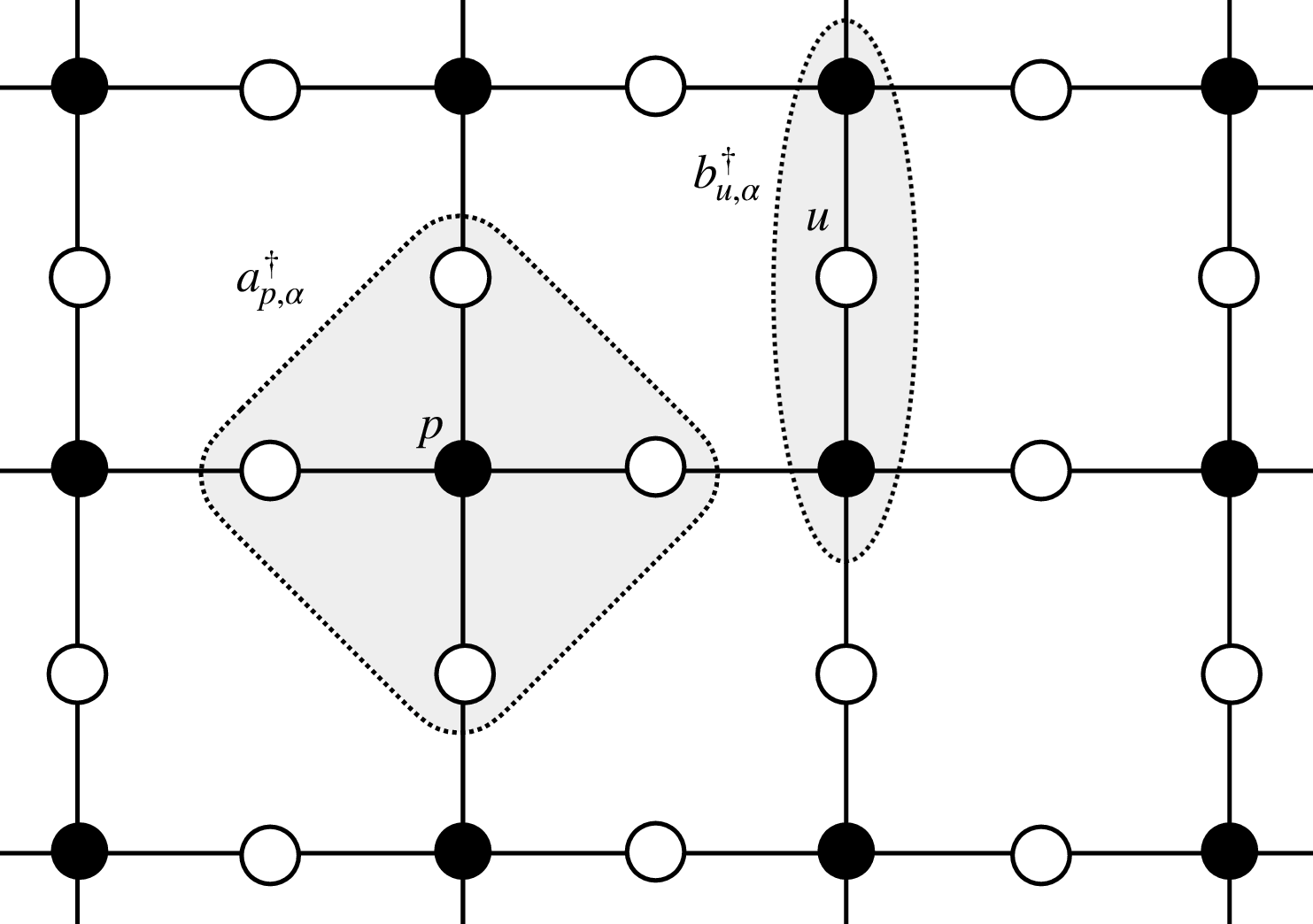}
	\caption{The lattice structure of $\Lambda$ for $d=2$.
	Black and white dots represent external sites in $\mathcal{E}$ and internal sites in $\mathcal{I}$, respectively.
	We also depict the localized states generated by $a_{p, \alpha}^{\dag}$ and $b_{u, \alpha}^{\dag}$.
	}
	\label{fig:lattice}       
\end{figure*}

Let $\mathcal{E}$ be a set of sites in a $d$-dimensional hypercubic lattice of length $L$ with unit lattice spacing and periodic boundary conditions, where we assume $d \ge 2$  and $L$ is an odd integer.
We take a new site in the middle of each bond of the lattice $\mathcal{E}$ and denote by $\mathcal{I}$ the collection of all such sites.
In the following, we call $p \in \mathcal{E}$ an external site and $u \in \mathcal{I}$ an internal site.
We define the SU($n$) Hubbard model on the decorated hypercubic lattice $\Lambda = \mathcal{E} \cup \mathcal{I}$.
The lattice structure for $d=2$ is shown in Fig.~\ref{fig:lattice}.


\subsection{Fermion Operatores} \label{subsec:fermion operators}
We denote creation and annihilation operators by $c_{x, \alpha}^{\dag}$ and $c_{x, \alpha}$ for a fermion at site $x \in \Lambda$ with color $\alpha = 1, \dots, n$.
They satisfy the anticommutation relations
\begin{align}
\{c_{x, \alpha}, c_{y, \beta}\} = \{c_{x, \alpha}^{\dag}, c_{y, \beta}^{\dag}\} = 0
\end{align}
and
\begin{align}
\{c_{x, \alpha}, c_{y, \beta}^{\dag}\} = \delta_{\alpha, \beta} \delta_{x, y}
\end{align}
for $x, y \in \Lambda$ and $\alpha, \beta = 1, \dots, n$.
The corresponding number operator is defined by $n_{x, \alpha} = c_{x, \alpha}^{\dag} c_{x, \alpha}$.
The total fermion number is $N_{\mathrm{f}} = \sum_{x \in \Lambda} n_{x} =\sum_{\alpha=1}^{n} \sum_{x \in \Lambda} n_{x, \alpha}$, where $n_{x} = \sum_{\alpha=1}^{n} n_{x, \alpha}$.
In the following, we consider $N_{\mathrm{f}}$-particle Hilbert space $\mathcal{H}_{N_{\mathrm{f}}}(\Lambda)$ with a fixed fermion number $N_{\mathrm{f}} = |\mathcal{E}| = L^{d}$, which means that the lowest band is 1/$n$ filled.
\par
We define color raising and lowering operators as 
\begin{align}
F^{\alpha, \beta} = \sum_{x \in \Lambda} c_{x, \alpha}^{\dag} c_{x, \beta} \ \ \text{for} \ \ \alpha \neq \beta
\end{align}
and total number operators of fermion with color $\alpha$ as
\begin{align}
F^{\alpha, \alpha} = \sum_{x \in \Lambda} c_{x, \alpha}^{\dag} c_{x, \alpha} \ \ \text{for} \ \ \alpha = 1, \dots, n.
\end{align}
We denote the eigenvalue of $F^{\alpha, \alpha}$ as $N_{\alpha}$.

To describe our model, we define a new set of operators
\begin{align}
a_{p, \alpha} &= c_{p, \alpha} - \nu \sum_{\substack{u \in \mathcal{I} \\ |p-u| = 1/2}} c_{u, \alpha} \ \ \text{for} \ \ p \in \mathcal{E},\\
b_{u, \alpha} &= c_{u, \alpha} + \nu \sum_{\substack{p \in \mathcal{E} \\ |p-u| = 1/2}} c_{p, \alpha} \ \ \text{for} \ \ u \in \mathcal{I},
\end{align}
where $\nu > 0$.
It is verified that these operators satisfy the following anticommutation relations for $p, q \in \mathcal{E}$ and $u, v \in \mathcal{I}$,
\begin{align}
\{a_{p, \alpha}, a_{q, \beta}\} 
&= \{b_{u, \alpha}, b_{v, \beta}\} = \{a_{p, \alpha}, b_{u, \beta}\} = 0, \label{eq:ac1}\\
\{a_{p, \alpha}, a_{q, \beta}^{\dag}\}
&= \begin{cases}
\delta_{\alpha, \beta} (2d\nu^{2} + 1) &\ \ \text{if} \ \ p=q, \\
\delta_{\alpha, \beta} \nu^{2} &\ \ \text{if} \ \ |p-q|=1, \\
0 &\ \ \text{otherwise}, 
\end{cases} \label{eq:ac2}\\
\{b_{u, \alpha}, b_{v, \beta}^{\dag}\} 
&= \begin{cases}
\delta_{\alpha, \beta}(2\nu^{2} + 1) &\ \ \text{if} \ \ u=v, \\
\delta_{\alpha, \beta} \nu^{2} &\ \ \text{if} \ \ u\neq v \ \text{and} \ C_{u} \cap C_{v} \neq \emptyset, \\
0 &\ \ \text{otherwise},
\end{cases} \label{eq:ac3}\\
\{a_{p, \alpha}, b_{u, \beta}^{\dag}\} &= 0, \label{eq:ac4}
\end{align}
where $C_{u}$ is a set of lattice sites consisting of $u$ itself and $p \in \mathcal{E}$ such that $|p-u| = 1/2$.
\par
\subsection{Model} \label{subsec:model}
First, we study the following SU($n$) Hubbard model
\begin{align}
H_{1} &= H_{\mathrm{hop}} + H_{\mathrm{int}}, \label{eq:ham1}\\
H_{\mathrm{hop}} &= t \sum_{\alpha=1}^{n} \sum_{u \in \mathcal{I}} b_{u, \alpha}^{\dag} b_{u, \alpha}, \label{eq:hop1}\\
H_{\mathrm{int}} &= U \sum_{\alpha < \beta} \sum_{x \in \Lambda} n_{x, \alpha} n_{x, \beta}, \label{eq:int}
\end{align}
where the parameters $t$, $U$ are non-negative.
When $U=0$, the model reduces to the tight-binding model described only by $H_{\mathrm{hop}}$.
The hopping Hamiltonian $H_{\mathrm{hop}}$ can be written in the standard form 
\begin{align}
H_{\mathrm{hop}} =\sum_{\alpha=1}^{n} \sum_{x, y \in \Lambda} t_{x, y} c_{x, \alpha}^{\dag} c_{y, \alpha},
\end{align} 
with hopping matrix elements given by
\begin{align}
t_{x, y} = 
\begin{cases}
\nu t &\ \ \text{if} \  |x-y|=1/2, \\
\nu^{2} t &\ \ \text{if} \ x, y \in \mathcal{E} \ \text{and} \ |x-y|=1, \\
t &\ \ \text{if} \ x=y \in \mathcal{I}, \\
2d\nu^{2} t &\ \ \text{if} \ x=y \in \mathcal{E}, \\
0 &\ \ \text{otherwise}.
\end{cases}
\end{align}
\par
\begin{figure*}
\centering
\includegraphics[scale=0.5]{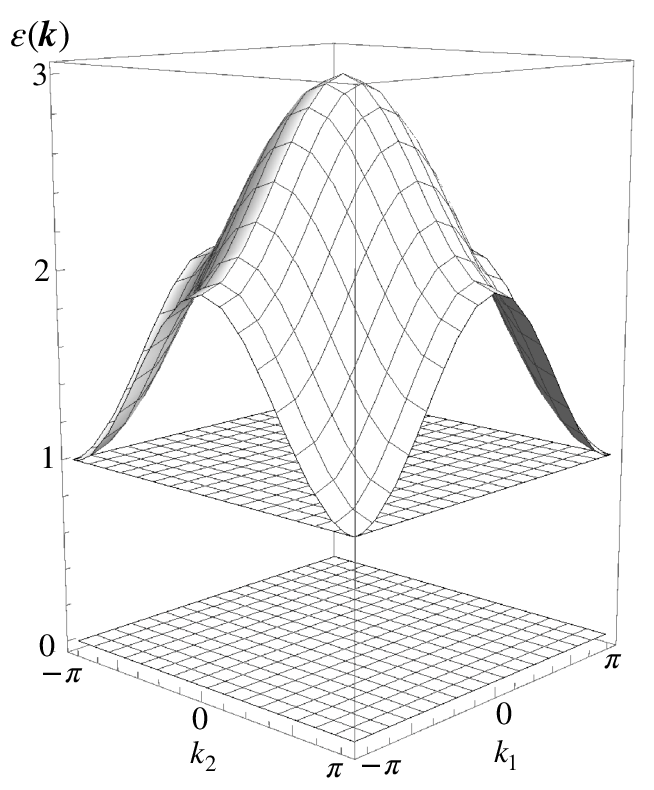}
\caption{The dispersion relations of the energy bands for $t=1$, $\nu=1/2$, and $d=2$.
 The lowest and middle bands are completely flat.}
\label{fig:flat band}       
\end{figure*}

Let us consider the corresponding single-particle Schr\"{o}dinger equation.
Let $\ket{\Phi_{1}}$ be a single-particle state of the form 
\begin{align}
\ket{\Phi_{1}} = \sum_{x \in \Lambda} \phi_{x} c_{x, \alpha}^{\dag} \ket{\Phi_{\mathrm{vac}}},
\end{align} 
where $\phi_{x} \in \mathbb{C}$ is a complex coefficient and $\ket{\Phi_{\mathrm{vac}}}$ is a vacuum state of $c_{x, \alpha}$.
One finds that the Schr\"{o}dinger equation $H_{1} \ket{\Phi_{1}} = \varepsilon \ket{\Phi_{1}}$ leads to
\begin{align}
\sum_{y \in \Lambda} t_{x,y} \phi_{y} = \varepsilon \phi_{x}, \label{eq:shreq1}
\end{align}
where $\varepsilon$ denotes a single-particle energy eigenvalue.
By solving Eq. (\ref{eq:shreq1}), one obtains the $d+1$ bands with dispersion relations 
\begin{align}
\varepsilon_{\mu}(\bm{k})
= \begin{cases}
0\ \ &\mu = 1,\\
t \ \ &\mu = 2, \dots, d, \\
t + 2t\nu^{2} \sum_{j=1}^{d}(1 + \cos{k_{j}}) \ \ &\mu = d+1,
\end{cases}
\end{align}
where $\bm{k}$ is an element in $\mathcal{K}$ defined by
\begin{align}
\mathcal{K} = \left\{\bm{k} = (k_{1}, \dots, k_{d}) \left| \ k_{j} = \frac{2\pi}{L} n_{j}, n_{j}= 0, \pm1, \dots, \pm \frac{L-1}{2} \ \text{for} \ j=1, \dots, d \right. \right\}. \label{eq:k set}
\end{align}
One finds that the lowest band and the middle bands are dispersionless, which are referred to as flat bands. 
See Fig.~\ref{fig:flat band} for the dispersion relations for $d=2$.
\par
\begin{figure*}
	\centering
	\includegraphics[scale=0.5]{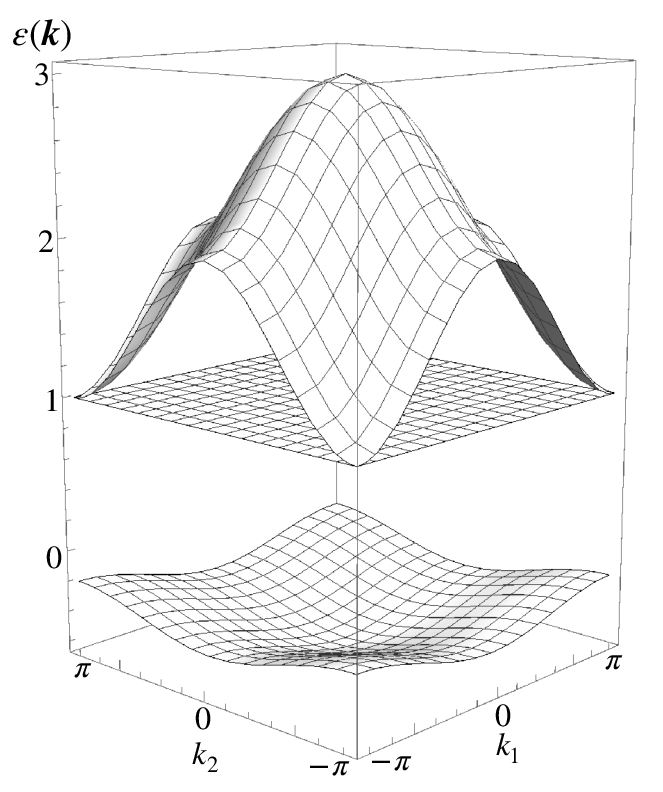}
	\caption{
	The dispersion relations of the energy bands for $t=1$, $s=1/5$, $\nu = 1/2$, and $d=2$.
	The lowest band is dispersive, while the middle band is still dispersionless.}
	\label{fig:nearly flat band}       
\end{figure*}

In the present paper, we also study a perturbed model, whose Hamiltonian is given by
\begin{align}
H_{2} &= H_{\mathrm{hop}}' + H_{\mathrm{int}}, \label{eq:ham2}\\
H_{\mathrm{hop}}' &=-s \sum_{\alpha=1}^{n} \sum_{p \in \mathcal{E}} a_{p, \alpha}^{\dag} a_{p, \alpha} + t \sum_{\alpha=1}^{n} \sum_{u \in \mathcal{I}} b_{u, \alpha}^{\dag} b_{u, \alpha}, \label{eq:hop2}
\end{align}
where the parameter $s$ and $t$ are non-negative and $H_{\mathrm{int}}$ is the same as Eq. (\ref{eq:int}).
The Hamiltonian $H_{2}$ is precisely the same as $H_{1}$ at $s=0$.
The hopping Hamiltonian $H_{\mathrm{hop}}'$ can also be written in the standard form
\begin{align}
H_{\mathrm{hop}}' = \sum_{x, y \in \Lambda} t_{x,y}' c_{x, \alpha}^{\dag} c_{y, \alpha},
\end{align}
where the hopping matrix elements are given by 
\begin{align}
t_{x,y}'
= \begin{cases}
\nu(t + s) & \ \ \text{if} \ |x-y| = 1/2, \\
\nu^{2} t & \ \ \text{if} \ x, y \in \mathcal{E} \ \text{and} \ |x-y|=1, \\
-\nu^{2} s & \ \ \text{if} \ x, y \in \mathcal{I} \ \text{and} \ x\neq y \ \text{and} \ ^{\exists}p \in \mathcal{E} \ \text{s.t.}\ |x-p|=|y-p|=1/2, \\
t - 2\nu^{2} s & \ \ \text{if} \ x=y \in \mathcal{I}, \\
2dt\nu^{2} - s & \ \ \text{if} \ x=y \in \mathcal{E}, \\
0 & \ \ \text{otherwise}.
\end{cases}
\end{align}
The corresponding single-particle Schr\"{o}dinger equation reads
\begin{align}
\sum_{y \in \Lambda} t_{x,y}' \phi_{y} = \varepsilon \phi_{x}. \label{eq:shreq2}
\end{align}
By solving Eq. (\ref{eq:shreq2}), one obtains the dispersion relations of $d+1$ bands
\begin{align}
\varepsilon_{\mu}(\bm{k})
= \begin{cases}
-s -2s \nu^{2} \sum_{j=1}^{d}(1 + \cos{k_{j}})\ \ &\mu=1,  \\
t \ \ &\mu=2, \dots, d,  \\
t + 2t\nu^{2} \sum_{j=1}^{d}(1 + \cos{k_{j}}) \ \ & \mu=d+1, 
\end{cases}
\end{align}
where $\bm{k}$ is an element of $\mathcal{K}$ defined in Eq. (\ref{eq:k set}).
Although the middle bands are still dispersionless, the lowest band has become dispersive because of the additional hopping term proportional to the parameter $s$.
See Fig.~\ref{fig:nearly flat band}.
 We note that all the single-particle properties of the hopping Hamiltonian are exactly the same as in the SU(2) case~\cite{tasaki1998nagaoka}.
\par
Since both of the Hamiltonians $H_{1}$ and $H_{2}$ have SU($n$) symmetry, the operators $F^{\alpha, \beta}$ commute with $H_{1}$ and $H_{2}$.
Therefore, the Hilbert space can be separated into different sectors labeled by $(N_{1}, \dots, N_{n})$, which is denoted by $\mathcal{H}_{N_{1}, \dots, N_{n}}(\Lambda)$. 
To establish our theorem, we define fully polarized states.
A fully polarized state with color $\alpha$ is defined 
as 
\begin{align}
\ket{\Phi_{\mathrm{all}, \alpha}} = \prod_{p \in \mathcal{E}} a_{p, \alpha}^{\dag} \ket{\Phi_{\mathrm{vac}}}, \label{eq:fully polarized state}
\end{align}
where $\ket{\Phi_{\mathrm{vac}}}$ is a vacuum state of $c_{x, \alpha}$.
From the anticommutation relations (\ref{eq:ac2}) and (\ref{eq:ac3}), we can see that the fully polarized state is an eigenstate of both $H_{1}$ and $H_{2}$.
Due to the SU($n$) symmetry, one obtains a general form of the eigenstates with the same energy as $\ket{\Phi_{\mathrm{all}, \alpha}}$:
\begin{align}
\ket{\Phi_{N_{1}, \dots, N_{n}}} = \left(F^{n, 1}\right)^{N_{n}} \cdots \left(F^{2, 1}\right)^{N_{2}} \ket{\Phi_{\mathrm{all}, 1}}, \label{eq:fully polarized state2}
\end{align}
where $N_{1} = |\mathcal{E}| - \sum_{\alpha=2}^{n} N_{\alpha}$.
We also refer to states of the form (\ref{eq:fully polarized state2}) as fully polarized states.
The fully polarized states can be characterized as eigenstates of the quadratic Casimir operator $C_{2}$ of the SU($n$) group defined as~\cite{ping2002group}
\begin{align}
C_{2} = \frac{1}{2} \left(\sum_{\alpha, \beta = 1}^{n} F^{\alpha, \beta} F^{\beta, \alpha} - \frac{N_{\rm f}^{2}}{n} \right).
\end{align}
The fully polarized states (\ref{eq:fully polarized state}) and (\ref{eq:fully polarized state2}) are eigenstates of $C_{2}$ with eigenvalue $\frac{|\mathcal{E}| (n-1)}{2} \left(\frac{|\mathcal{E}|}{n} + 1\right)$, which is the maximum eigenvalue of $C_{2}$ for fixed $N_{\rm f}$.
\subsection{Results} \label{subsec:results}
First, we describe the theorem for the model whose lowest band is flat.
This is a slight generalization of the result obtained by Liu {\it et al.}, in~\cite{liu2019flat}, in the sense that our hopping Hamiltonian has one more parameter.
{\theorem \label{thm:1}
Consider the SU($n$) Hubbard Hamiltonian (\ref{eq:ham1}) with the total fermion number $N_{\mathrm{f}} = |\mathcal{E}|$.
For arbitrary $t > 0$ and $U > 0$, the ground states of the Hamiltonian are the fully polarized states and unique apart from trivial degeneracy due to the SU($n$) symmetry.} 
\par
The theorem is proved in Sect.~\ref{sec:3}.
The fully polarized states are the SU($n$) counterparts of the ferromagnetic states in the SU(2) Hubbard model.
Thus, Theorem~\ref{thm:1} establishes the SU($n$) ferromagnetism in the $d$-dimensional Hubbard models with flat bands.
\par
Since the density of states at the Fermi energy diverges, Theorem~\ref{thm:1} can be thought of as a result for a singular case.
In order to establish a rigorous result for a nonsingular SU($n$) Hubbard model, we consider the Hamiltonian (\ref{eq:ham2}).
To state the theorem, let us define for $n\ge 2$ and $d \ge 2$, 
\begin{align}
\nu_{\mathrm{c}}(n, d)
=\begin{cases}
\sqrt{\frac{2d+1 + \sqrt{4(2d-n) (n-1) + (2d+1)^{2}}}{2(2d-n) (n-1)}}  &\ \ \text{for} \ \ n \le d, \\
\sqrt{\frac{2d+1 + \sqrt{8d^{2} + 1}}{2d(d-1)}} &\ \ \text{for} \ \ n > d.
\end{cases}
\end{align}

{\theorem \label{thm:2}
Consider the SU($n$) Hubbard Hamiltonian (\ref{eq:ham2}) with the total fermion number $N_{\mathrm{f}} = |\mathcal{E}|$.
Assume that $0 < \nu \le \nu_{\mathrm{c}}(n, d)$ for $d \ge 2$.
For sufficiently large $t/s>0$ and $U/s>0$, the ground states are the fully polarized states and unique apart from the trivial degeneracy due to the SU($n$) symmetry.} 
\par
Since the lowest band of Eq. (\ref{eq:hop2}) is dispersive, the density of states at the Fermi level does not diverge.
Also, the Coulomb interaction is assumed to be sufficiently large but finite.
Thus, the theorem establishes SU($n$) ferromagnetism in the ground states of non-singular SU($n$) Hubbard models in arbitrary dimensions.
The theorem is proved in Sect.~\ref{sec:4}.
\section{Proof of Theorem~\ref{thm:1}\label{sec:3}} 
We shall prove Theorem~\ref{thm:1} in this section.
We consider the model (\ref{eq:ham1}) with the fermion number $N_{\mathrm{f}} = |\mathcal{E}|$.
First, we note that the hopping Hamiltonian $H_{\mathrm{hop}}$ and $H_{\mathrm{int}}$ are positive semidefinite because $b_{u, \alpha}^{\dag} b_{u, \alpha}$ and $n_{x, \alpha} n_{x, \beta} = (c_{x, \alpha} c_{x, \beta})^{\dag} c_{x, \alpha} c_{x, \beta}$ are positive semidefinite.
Hence, the total Hamiltonian $H_{1}$ is positive semidefinite as well.
By noting $\{a_{p, \alpha}, b_{u, \beta}^{\dag}\} = 0$, we find that the fully polarized state $\ket{\Phi_{\mathrm{all}, \alpha}}$ is an eigenstate of $H_{1}$ with eigenvalue zero and all the fully polarized states are zero energy states due to the SU($n$) symmetry.
Since $H_{1} \ge 0$, the fully polarized states are ground states of $H_{1}$.
\par
In the following, we prove the uniqueness of ground states.
Let $\ket{\Phi_{\mathrm{GS}}}$ be an arbitrary ground state of $H_{1}$ with $N_{\mathrm{f}} = |\mathcal{E}|$, which means $H_{1} \ket{\Phi_{\mathrm{GS}}} = 0$.
We note that in general, the ground state can be written as 
\begin{align}
&\ket{\Phi_{\mathrm{GS}}} = \nonumber\\ 
&\sum_{\substack{A_{1}, \dots, A_{n} \subset \mathcal{E}\\
B_{1}, \dots, B_{n} \subset \mathcal{I} \\
}} f(\{ A_{\alpha}\}, \{B_{\alpha}\}) \left(\prod_{p \in A_{1}} a_{p, 1}^{\dag}\right) \cdots \left(\prod_{p \in A_{n}} a_{p, n}^{\dag}\right) \left(\prod_{u \in B_{1}} b_{u, 1}^{\dag}\right) \cdots \left(\prod_{u \in B_{n}} b_{u, n}^{\dag}\right) \ket{\Phi_{\mathrm{vac}}},
\end{align}
where $A_{\alpha}$ and $B_{\alpha}$ are subsets of $\mathcal{E}$ and $\mathcal{I}$, respectively such that $\sum_{\alpha=1}^{n} \left(|A_{\alpha}| + |B_{\alpha}|\right) = |\mathcal{E}|$.
Here, $f(\{A_{\alpha}\}, \{B_{\alpha}\})$ is a certain coefficient.
The inequalities $H_{\mathrm{hop}} \ge 0$ and $H_{\mathrm{int}} \ge 0$ imply that $H_{\mathrm{hop}} \ket{\Phi_{\mathrm{GS}}} = 0 $ and $H_{\mathrm{int}} \ket{\Phi_{\mathrm{GS}}} = 0$, which yield
\begin{align}
b_{u, \alpha} \ket{\Phi_{\mathrm{GS}}} &= 0 \ \text{for any} \ u \in \mathcal{I} \ \text{and} \ \ \alpha = 1, \dots, n, \label{eq:condition1}\\
c_{x, \alpha} c_{x, \beta} \ket{\Phi_{\mathrm{GS}}} &= 0 \ \text{for any} \ x \in \Lambda \ \text{and} \ \ \alpha \neq \beta. \label{eq:condition2}
\end{align}
From the anticommutation relations (\ref{eq:ac4}), it is clear that the conditions (\ref{eq:condition1}) imply that $\ket{\Phi_{\mathrm{GS}}}$ does not contain any $b_{u, \alpha}^{\dag}$ operator.
(See Appendix~\ref{app:1} for a proof that the states obtained by acting with $a^{\dag}$ and $b^{\dag}$ operators on $\ket{\Phi_{\mathrm{vac}}}$ are linearly independent.
\footnote{See~\cite{mielketasaki1993ferromagnetism} for a detailed discussion.}
Therefore, the ground state is written as 
\begin{align}
\ket{\Phi_{\mathrm{GS}}} 
= \sum_{\substack{A_{1} , \dots, A_{n} \subset \mathcal{E} \\
\sum_{\alpha=1}^{n} |A_{\alpha}| = |\mathcal{E}|}
}
g\left(\{A_{\alpha}\}\right) \left(\prod_{p \in A_{1}} a_{p, 1}^{\dag}\right) \cdots  \left(\prod_{p \in A_{n}} a_{p, n}^{\dag}\right) \ket{\Phi_{\mathrm{vac}}},
\end{align}
where $g\left(\{A_{\alpha}\}\right)$ is a certain coefficient. 
\par
We next examine the conditions (\ref{eq:condition2}).
We first consider the case where $x=p \in \mathcal{E}$.
It follows from
\begin{align}
\{c_{p, \alpha}, a_{q, \beta}^{\dag}\} = \delta_{\alpha, \beta} \delta_{p, q} \ \ \text{for} \ p, q \in \mathcal{E}. \label{eq:ac c and a}
\end{align}
and Eq. (\ref{eq:condition2}) that $g(\{A_{\alpha}\}) = 0$ if there exist $A_{\alpha}$ and $A_{\beta}$ such that $A_{\alpha} \cap A_{\beta} \neq \emptyset$.
Noting that $\sum_{\alpha=1}^{n} |A_{\alpha}| = |\mathcal{E}|$, we find that $\cup_{\alpha=1}^{n} A_{\alpha} = \mathcal{E}$ when $A_{\alpha} \cap A_{\beta} = \emptyset$ for all $\alpha \neq \beta$.
Thus one can rewrite the ground state in the form of
\begin{align}
\ket{\Phi_{\mathrm{GS}}} = \sum_{\boldsymbol{\alpha}} C(\boldsymbol{\alpha}) \left( \prod_{p \in \mathcal{E}} a_{p, \alpha_{p}}^{\dag}\right) \ket{\Phi_{\mathrm{vac}}}, \label{eq:a only}
\end{align}
where the sum is over all possible color configuration $\boldsymbol{\alpha} = (\alpha_{p})_{p \in \mathcal{E}}$ with $\alpha_{p} = 1, \dots, n$.
In the derivation of Eq. (\ref{eq:a only}), we again exploit the linear independence of the states consisting of $a^{\dag}$ and $b^{\dag}$ operators, which is proved in Appendix \ref{app:1}.
Then we consider the conditions (\ref{eq:condition2}) for $x=u \in \mathcal{I}$.
For $u \in \mathcal{I}$, we can check that
\begin{align}
\{c_{u, \alpha}, a_{p, \beta}^{\dag}\}
=\begin{cases}
-\nu \delta_{\alpha, \beta} \ \ &\text{if} \ |u-p| = 1/2, \\
0 \ \ &\text{otherwise}.
\end{cases} \label{eq:ac c and a 2}
\end{align}
Using Eq. (\ref{eq:ac c and a 2}), we obtain 
\begin{align}
c_{u, \alpha} c_{u, \beta} \ket{\Phi_{\mathrm{GS}}}
= \sum_{\substack{
\boldsymbol{\alpha} \\
\mathrm{s.t.} \alpha_{p} = \beta, \alpha_{q} = \alpha
}}
(C(\boldsymbol{\alpha}) - C(\boldsymbol{\alpha}_{p \leftrightarrow q})) \left(\prod_{p' \in \mathcal{E} \backslash \{p, q\}}a_{p', \alpha_{p'}}^{\dag} \right) \ket{\Phi_{\mathrm{vac}}}, 
\end{align}
where $p$ and $q$ are external sites which satisfy $|u-p| = |u-q| = 1/2$.
The color configuration $\boldsymbol{\alpha}_{p \leftrightarrow q}$ is obtained from $\boldsymbol{\alpha}$ by swapping $\alpha_{p}$ and $\alpha_{q}$.
Since all the states in the sum are linearly independent (see Appendix~\ref{app:1}), we find from Eq. (\ref{eq:condition2}) that $C(\boldsymbol{\alpha}) = C(\boldsymbol{\alpha}_{p \leftrightarrow q})$ for all $\boldsymbol{\alpha}$ and all $u \in \mathcal{I}$.
Recalling that there is one localized state centered at each external site and neighboring localized states always share one internal site, we see that $C(\boldsymbol{\alpha}) = C(\boldsymbol{\alpha}_{p \leftrightarrow q})$ for $p$, $q \in \mathcal{E}$. 
Since an arbitrary permutation of the color configuration $\bm{\alpha}$ can be generated by repeatedly swapping the colors on neighboring external sites, we have
\begin{align}
C(\bm{\alpha}) = C(\bm{\beta}) \label{eq:fully symmetric}
\end{align}
if $\bm{\beta}$ is a permutation of $\bm{\alpha}$.
\par
Lastly, we show that the states satisfying Eq. (\ref{eq:fully symmetric}) are the fully polarized states defined in Eqs. (\ref{eq:fully polarized state}), (\ref{eq:fully polarized state2}).
To this aim, we introduce a concept of a word~\cite{kitaev2011patterns}.
A word $w = (w_{1}, \dots, w_{|\mathcal{E}|})$ is a sequence of integers where $w_{i} \in \{1, \dots, n\}$ for all $i$.
We denote by $|w|_{\alpha}$ the number of occurrences of $\alpha$ in $w$.
We define the set of words for which $|w|_{\alpha} = N_{\alpha}$ holds as follows: $W(N_{1}, \dots, N_{n}) = \{w| \ |w|_{\alpha} = N_{\alpha}, \alpha=1, \dots, n\}$.
For example, $W(2, 0, 1)$ consists of $(1, 1, 3)$, $(1, 3, 1)$ and $(3, 1, 1)$.
It follows from Eq. (\ref{eq:fully symmetric}) that the ground state of $H_{1}$ in the sector $\mathcal{H}_{N_{1}, \dots, N_{n}}(\Lambda)$ can be written as 
\begin{align}
\ket{\tilde{\Phi}_{N_{1}, \dots, N_{n}}} = \sum_{w \in W(N_{1}, \dots, N_{n})} a_{p_{1}, w_{1}}^{\dag} a_{p_{2}, w_{2}}^{\dag} \cdots a_{p_{|\mathcal{E}|}, w_{|\mathcal{E}|}}^{\dag} \ket{\Phi_{\mathrm{vac}}},
\end{align}
where each external site is labeled by an integer $i$ as $p_{i}$.
On the other hand, using commutation relations $[F^{\beta, \alpha}, a_{p, \gamma}^{\dag}] = \delta_{\alpha, \gamma} a_{p, \beta}^{\dag}$, we see that
\begin{align}
\left(F^{2, 1}\right)^{N_{2}} \ket{\Phi_{\mathrm{all}, 1}}
= N_{2}! \sum_{w \in W(|\mathcal{E}|-N_{2}, N_{2})} a_{p_{1}, w_{1}}^{\dag} a_{p_{2}, w_{2}}^{\dag} \cdots a_{p_{|\mathcal{E}|}, w_{|\mathcal{E}|}}^{\dag} \ket{\Phi_{\mathrm{vac}}}.
\end{align}
By repeating the same procedure, we have the desired result, i.e., $\ket{\Phi_{N_{1}, \dots, N_{n}}}$ and $\ket{\tilde{\Phi}_{N_{1}, \dots, N_{n}}}$ are the same up to a normalization.
Thus, Theorem~\ref{thm:1} is proved.
\hspace{\fill}$\square$
\section{Proof of Theorem~\ref{thm:2}\label{sec:4}} 
In this section, we shall prove Theorem~\ref{thm:2}.
Here, we consider the model (\ref{eq:ham2}) with the fermion number $N_{\mathrm{f}} = |\mathcal{E}|$.
First, we rewrite the Hamiltonian $H_{2}$ as follows:
\begin{align}
H_{2} = -s |\mathcal{E}| (2d\nu^{2} + 1) + \lambda H_{\mathrm{flat}} + \sum_{p \in \mathcal{E}} h_{p},
\end{align}
where 
\begin{align}
H_{\mathrm{flat}} = \sum_{\alpha=1}^{n} \sum_{u \in \mathcal{I}} b_{\alpha, u}^{\dag} b_{\alpha, u} 
+ \sum_{\alpha < \beta} \sum_{x\in \Lambda} n_{x, \alpha} n_{x, \beta},
\end{align}
and the local Hamiltonian for each $p \in \mathcal{E}$ is defined as
\begin{align}
&h_{p} 
= s(2d\nu^{2} + 1) - s \sum_{\alpha=1}^{n} a_{p, \alpha}^{\dag} a_{p, \alpha} + \frac{t-\lambda}{2} \sum_{\alpha=1}^{n} \sum_{\substack{u \in \mathcal{I} \\ |p-u|=1/2}} b_{u, \alpha}^{\dag} b_{u, \alpha} \nonumber  \\
& + \sum_{\alpha < \beta} \left[
(U-\lambda)(1-\kappa) n_{p, \alpha} n_{p, \beta} 
+ \frac{\kappa(U-\lambda)}{2d} \sum_{\substack{q \in \mathcal{E} \\ |p-q|=1}} n_{q, \alpha} n_{q, \beta}
+ \frac{U -\lambda}{2} \sum_{\substack{u \in \lambda \\ |p-u|=1/2}} n_{u, \alpha} n_{u, \beta}
\right].
\end{align}
The two parameters $\lambda$ and $\kappa$ satisfy $0 < \lambda < \min\{t, U\}$ and $0 < \kappa < 1$.
To prove Theorem~\ref{thm:2}, we prove the following lemma.
{\lemma \label{lemma:1}
Suppose the local Hamiltonian $h_{p}$ is positive semidefinite for any $p \in \mathcal{E}$.
Then the ground states of the Hamiltonian $H_{2}$ are fully polarized states and unique apart from the trivial degeneracy due to the SU($n$) symmetry.
}
{\proof
First, we can check that a fully polarized state $\ket{\Phi_{\mathrm{all}, 1}}$ satisfies $h_{p} \ket{\Phi_{\mathrm{all}, 1}} = 0$ for all $p \in \mathcal{E}$.
Since the local Hamiltonian is SU($n$) symmetric, all the fully polarized states have zero energy for $h_{p}$.
From the assumption that $h_{p}$ is positive semidefinite, it follows that the fully polarized states are the ground states of $h_{p}$ for all $p\in \mathcal{E}$.
These states are also the ground states of $H_{\mathrm{flat}}$ since $H_{\mathrm{flat}}$ is equal to $H_{1}$ with $t=U=1$.
Hence, the fully polarized states are the ground states of $H_{2}$ with energy $-s|\mathcal{E}|(2d\nu^{2} + 1)$.
On the other hand, any ground state of $H_{2}$ must be a simultaneous ground state of $h_{p}$ and $H_{\mathrm{flat}}$ since both $h_{p}$ and $H_{\mathrm{flat}}$ are positive semidefinite.
It is shown from Theorem~\ref{thm:1} that the ground states of $H_{\mathrm{flat}}$ are fully polarized states and unique apart from trivial degeneracy due to the SU($n$) symmetry.
Thus, Lemma~\ref{lemma:1} is proved.
\hspace{\fill} $\square$
}
\par
We can establish the positive semidefiniteness of $h_{p}$ by the following lemma.
{\lemma \label{lemma:2}
Suppose that $t/s, U/s$ are sufficiently large and $0 < \nu < \nu_{c}(n, d)$.
Then the local Hamiltonian $h_{p}$ is positive semidefinite. 
(We take $\lambda$ and $\kappa$ to be proportional to $s$.)
}
\par
{\proof
Because of the translational invariance, it suffices to prove the lemma for $h_{o}$ with a fixed $o \in \mathcal{E}$.
For convenience, we set $o = (0, 0, \dots, 0) \in \mathbb{Z}^{d}$.
We regard $h_{o}$ as an operator acting on a lattice $\Lambda_{o}$ with $4d+1$ sites.
This lattice is written as $\Lambda_{o} = \{o\} \cup \mathcal{E}_{o} \cup \mathcal{I}_{o}$, where $\mathcal{E}_{o}$ is a set of $2d$ external sites of the form $(0, \dots, 0, \pm 1 ,0,  \dots, 0)$, and $\mathcal{I}_{o}$ is a set of $2d$ internal sites of the form $(0, \dots, 0, \pm 1/2, 0, \dots, 0)$.
See Fig.~\ref{fig:Lambdao}.
\begin{figure*}
	\begin{tabular}{c}
		\begin{minipage}{0.5 \hsize}
		\subcaption{}\label{fig:Lambdao a}
		\centering
		\includegraphics[scale=0.35]{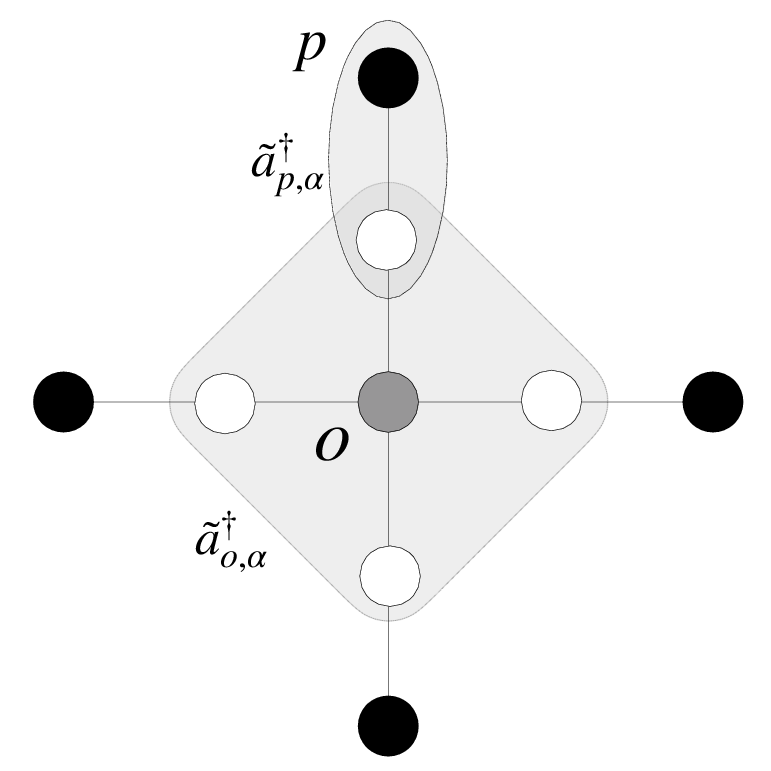}
		\end{minipage}
	
		\begin{minipage}{0.5 \hsize}
		\subcaption{}\label{fig:Lambdao b}
		\centering
		\includegraphics[scale=0.35]{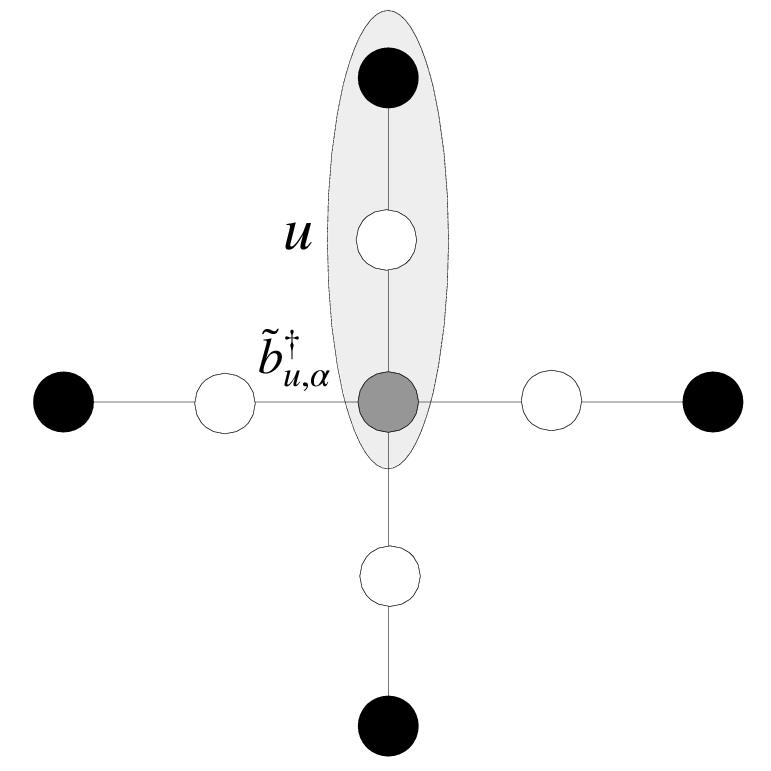}
		\end{minipage}	
	\end{tabular}
	\caption{The lattice $\Lambda_{o}$ with $4d+1$ sites for $d=2$.
	The origin $o$ is located at the center.
	The sets $\mathcal{E}_{o}$ and $\mathcal{I}_{o}$ consist of black dots and white dots, respectively.
	We draw \subref{fig:Lambdao a} the states generated by $\tilde{a}_{o, \alpha}^{\dag}$ and $\tilde{a}_{p, \alpha}^{\dag}$ and \subref{fig:Lambdao b} the state generated by $\tilde{b}_{u, \alpha}^{\dag}$.
	}
	\label{fig:Lambdao}       
\end{figure*}
In the following, we denote by $\mathcal{F}(\Lambda_{o})$ the Fock space on $\Lambda_{o}$.
We define the total fermion number and the total number operator of fermion with color $\alpha$ on $\mathcal{F}(\Lambda_{o})$ as
\begin{align}
F = \sum_{x \in \Lambda_{o}} n_{x}, 
\end{align}
and 
\begin{align}
M_{\alpha} = \sum_{x \in \Lambda_{o}} n_{x, \alpha},
\end{align}
respectively.
\par
It is convenient to define operators on $\mathcal{F}(\Lambda_{o})$,
\begin{align}
\tilde{a}_{o, \alpha} &= c_{o, \alpha} - \nu \sum_{u \in \mathcal{I}_{o}} c_{u, \alpha}, \\
\tilde{a}_{p, \alpha} &= \frac{1}{\sqrt{\nu^{2} + 1}} \left(c_{p, \alpha} - \nu c_{p/2, \alpha} \right) \ \ \text{for} \ p \in \mathcal{E}_{o}, \\ 
\tilde{b}_{u, \alpha} &= c_{u, \alpha} + \nu c_{o, \alpha} + \nu c_{2u, \alpha} \ \ \text{for} \ u \in \mathcal{I}_{o}.
\end{align}
These operators satisfy the anticommutation relations
\begin{align}
\{\tilde{a}_{p, \alpha}, \tilde{a}_{q, \beta}^{\dag}\}
&=
\begin{cases}
\delta_{\alpha, \beta} (2d\nu^{2} + 1 ) \ \ &\text{for} \ p = q = o, \\
\delta_{\alpha, \beta} \frac{\nu^{2}}{\sqrt{\nu^{2} + 1}} \ \ &\text{for} \ p=o, q \in \mathcal{E}_{o}, \\
\delta_{\alpha, \beta} \delta_{p, q} \ \ &\text{for} \ p, q \in \mathcal{E}_{o},
\end{cases} \\
\{\tilde{b}_{u, \alpha}, \tilde{b}_{v, \beta}^{\dag}\} 
&=
\begin{cases}
\delta_{\alpha, \beta}(2\nu^{2} + 1) \ \ &\text{for} \ u=v, \\
\delta_{\alpha, \beta} \nu^{2} \ \ &\text{for} \ u \neq v,
\end{cases}\\
\{\tilde{a}_{p, \alpha}, \tilde{b}_{u, \beta}^{\dag}\} 
&= 0 \ \ \text{for} \ p \in \{o\} \cup \mathcal{E}_{o}, u \in \mathcal{I}_{o}.
\end{align}
With these operators, $h_{o}$ can be written as 
\begin{align}
&h_{o} = s(2d\nu^{2} + 1) - s \sum_{\alpha=1}^{n} \tilde{a}_{o, \alpha}^{\dag} \tilde{a}_{o, \alpha} 
+ \frac{t-\lambda}{2} \sum_{\alpha=1}^{n} \sum_{u\in \mathcal{I}_{o}} \tilde{b}_{u, \alpha}^{\dag} \tilde{b}_{u, \alpha} \nonumber \\
&+ \sum_{\alpha <\beta} \left[
(U-\lambda)(1-\kappa) n_{o, \alpha} n_{o, \beta} + \frac{\kappa(U-\lambda)}{2d} \sum_{p \in \mathcal{E}_{o}} n_{p, \alpha} n_{q, \alpha} + \frac{U-\lambda}{2} \sum_{u \in \mathcal{I}_{o}} n_{u, \alpha} n_{u, \beta}
\right].
\end{align}
\par
Since the total fermion number $F$ on $\Lambda_{o}$ is conserved, we can examine $h_{o}$ in each fermion-number sector separately.
The Fock space can be decomposed into a direct sum as
\begin{align}
\mathcal{F}(\Lambda_{o}) = \bigoplus_{F=0}^{n(4d+1)} \mathcal{H}_{F}(\Lambda_{o}),
\end{align}
where $\mathcal{H}_{F}(\Lambda_{o})$ is the $F$-fermion Hilbert space.
Note that the lattice $\Lambda_{o}$ can accommodate at most $n(4d+1)$ fermions.
Since $M_{\alpha}$ ($\alpha=1, \dots, n$) commutes with $h_{o}$, we can further decompose each fermion-number sector $\mathcal{H}_{F}(\Lambda_{o})$ into a direct sum of smaller subspaces as
\begin{align}
\mathcal{H}_{F}(\Lambda_{o}) = \bigoplus_{\substack{(M_{1}, \dots, M_{n}) \\ \sum_{\alpha=1}^{n} M_{\alpha} = F}} \mathcal{J}_{M_{1}, \dots, M_{n}}(\Lambda_{o}), \label{eq:J decomposition}
\end{align}
where $\mathcal{J}_{M_{1}, \dots, M_{n}}(\Lambda_{o})$ is the Hilbert space with each $M_{\alpha}$ fixed.
First, we consider the one-particle sector $\mathcal{H}_{1}(\Lambda_{o})$.
The noninteracting part of $h_{o}$ is 
\begin{align}
s(2d\nu^{2} + 1) - s \sum_{\alpha=1}^{n} \tilde{a}_{o, \alpha}^{\dag} \tilde{a}_{o, \alpha} + \frac{t-\lambda}{2} \sum_{\alpha=1}^{n} \sum_{u \in \mathcal{I}_{o}} \tilde{b}_{u, \alpha}^{\dag} \tilde{b}_{u, \alpha}, 
\end{align}
and one finds that the single-particle eigenenergies are $0$, $s(2d\nu^{2} + 1)$, $s(2d\nu^{2} + 1) + (t-\lambda)((2d+1)\nu^{2} + 1)/2$, $s(2d\nu^{2} + 1) + (t-\lambda)(\nu^{2} + 1)/2$.
These eigenenergies are non-negative.
Therefore, we see that the local ground state energy $E_{1}^{\mathrm{GS}}$ in $\mathcal{H}_{1}(\Lambda_{o})$ is given by
\begin{align}
E_{1}^{\mathrm{GS}} = 0,
\end{align}
and $h_{o}$ within this subspace is positive semidefinite.
\par
Next, we consider $\mathcal{H}_{F}(\Lambda_{o})$ where $ 2 \le F \le n(4d+1)$.
We shall prove 
\begin{align}
\lim_{t, U \rightarrow \infty} \mel{\Phi}{h_{o}}{\Phi} \geq 0\label{eq:p.s.d.}
\end{align}
for any normalized state $\ket{\Phi} \in \mathcal{H}_{F}(\Lambda_{o})$ for all $F$.
To prove Eq. (\ref{eq:p.s.d.}), we only need to examine states such that $\lim_{t, U \rightarrow \infty} \mel{\Phi}{h_{o}}{\Phi} < \infty$, which we call finite-energy states.
Since $\tilde{b}_{u, \alpha}^{\dag}\tilde{b}_{u, \alpha}$ and $n_{x, \alpha} n_{x, \beta}$ are positive semidefinite, the condition that $\ket{\Phi}$ is a finite-energy state is equivalent to the following:
\begin{align}
\tilde{b}_{u, \alpha} \ket{\Phi} &= 0 \ \ \text{for all} \ u\in \mathcal{I}_{o} \\
c_{x, \alpha} c_{x, \beta} \ket{\Phi} &= 0 \ \ \text{for all} \ x \in \Lambda_{o} \ \text{and} \ \alpha \neq \beta.
\end{align}
Repeating the same argument in the proof of Theorem~\ref{thm:1}, we find that the conditions $\tilde{b}_{u, \alpha} \ket{\Phi} = 0$ for any $u \in \mathcal{I}_{o}$ and $c_{p, \alpha} c_{p, \beta} \ket{\Phi} = 0$ for $p \in \mathcal{E}_{o}$ imply that any finite-energy state $\ket{\Phi}$ is in the form of 
\begin{align}
\ket{\Phi} 
= \sum_{\substack{
A_{1} ,\dots, A_{n} \subset \{o\} \cup \mathcal{E}_{o} \\
A_{\alpha} \cap A_{\beta} = \emptyset  \ \text{for all} \ \alpha \neq \beta \\
\sum_{\alpha=1}^{n} |A_{\alpha}| = F
}}  f' \left(\{A_{\alpha}\}\right) \left(\prod_{p \in A_{1}} \tilde{a}_{p, 1}^{\dag}\right) \dots \left(\prod_{p \in A_{n}} \tilde{a}_{p, n}^{\dag}\right) \ket{\Phi_{\mathrm{vac}}},
\end{align}
where $A_{\alpha}$ is a subset of $\{o\} \cup \mathcal{E}_{o}$ such that $A_{\alpha} \cap A_{\beta} = \emptyset$ for $\alpha \neq \beta$ and $\sum_{\alpha=1}^{n}|A_{\alpha}| = F$, and $f'(\{A_{\alpha}\})$ is a certain coefficient.
We divide the collections of subsets $\{A_{\alpha}\}$ into two groups: those that contain the origin $o$ and those that do not, and then we express the finite-energy state as
\begin{align}
\ket{\Phi} = \ket{\Phi_{o}} + \ket*{\tilde{\Phi}},
\end{align}
where 
\begin{align}
\ket{\Phi_{o}} 
&= \sum_{\substack{
A_{1}, \dots, A_{n} \in \{o\} \cup \mathcal{E}_{o} \\
A_{\alpha} \cap A_{\beta} = \emptyset \ \text{for all} \ \alpha \neq \beta \\
o \in \cup_{\alpha=1}^{n} A_{\alpha}, \sum_{\alpha=1}^{n}|A_{\alpha}| = F
}}
f' \left(\{A_{\alpha}\}\right) \left(\prod_{p \in A_{1}} \tilde{a}_{p, 1}^{\dag}\right) \dots \left(\prod_{p \in A_{n}} \tilde{a}_{p, n}^{\dag}\right) \ket{\Phi_{\mathrm{vac}}}, \label{eq:contain o}\\
\ket*{\tilde{\Phi}}
&=\sum_{\substack{
A_{1}, \dots, A_{n} \in \mathcal{E}_{o} \\
A_{\alpha} \cap A_{\beta} = \emptyset \ \text{for all} \ \alpha \neq \beta \\
\sum_{\alpha=1}^{n}|A_{\alpha}| = F
}}
f' \left(\{A_{\alpha}\}\right) \left(\prod_{p \in A_{1}}\tilde{a}_{p, 1}^{\dag}\right) \dots \left(\prod_{p \in A_{n}}\tilde{a}_{p, n}^{\dag}\right) \ket{\Phi_{\mathrm{vac}}}. \label{eq:not contain o}
\end{align}
Using the conditions $c_{u, \alpha} c_{u, \beta} \ket{\Phi} = 0$ for $u \in \mathcal{I}_{o}$, we see that $\ket{\Phi_{o}}$ should be a fully polarized state and $h_{o} \ket{\Phi_{o}} = 0$.
Therefore, the expectation value $\mel{\Phi}{h_{o}}{\Phi}$ is the same as $\mel*{\tilde{\Phi}}{h_{o}}{\tilde{\Phi}}$ and it is enough to examine normalized states in the form of Eq. (\ref{eq:not contain o}).
We can verify that for any states written as Eq. (\ref{eq:not contain o}) it holds that 
\begin{align}
\tilde{a}_{o, \alpha}^{\dag} \ket*{\tilde{\Phi}} = \frac{\nu^{2}}{\sqrt{\nu^{2} + 1}} \sum_{p \in \mathcal{E}_{o}} \tilde{a}_{p, \alpha}^{\dag} \ket*{\tilde{\Phi}},
\end{align}
and hence 
\begin{align}
\mel*{\tilde{\Phi}}{h_{o}}{\tilde{\Phi}}
= s(2d\nu^{2} + 1) - \frac{s\nu^{4}}{\nu^{2} + 1} \sum_{\alpha=1}^{n} \sum_{p, q \in \mathcal{E}_{o}}  \bra*{\tilde{\Phi}} \tilde{a}_{p, \alpha}^{\dag} \tilde{a}_{q, \alpha} \ket*{\tilde{\Phi}}. \label{eq:exp of ho}
\end{align}
Since the state $\ket*{\tilde{\Phi}}$ does not have any double occupancy of $\tilde{a}_{p, \alpha}$ states, to evaluate Eq. (\ref{eq:exp of ho}), it is sufficient to consider the effective Hamiltonian
\begin{align}
h_{\mathrm{eff}} = E_{0} P - s' \sum_{\alpha=1}^{n} \sum_{p, q \in \mathcal{E}_{o}} P \tilde{a}_{p, \alpha}^{\dag} \tilde{a}_{q, \alpha} P,
\end{align}
where $E_{0} = s(2d\nu^{2} + 1)$ and $s' = s\nu^{4}/(\nu^{2} + 1)$.
The operator $P = \prod_{\alpha < \beta} \prod_{x \in \mathcal{E}_{o}} \left(1 - \tilde{a}_{p, \alpha}^{\dag} \tilde{a}_{p, \alpha} \tilde{a}_{p, \beta}^{\dag} \tilde{a}_{p, \beta}\right)$ denotes the projection operator onto the space with no double occupancy of $\tilde{a}_{p, \alpha}$ states, which is known as the Gutzwiller projection~\cite{fazekas1999lecture}.
Here, we note that the operators $\tilde{a}_{p, \alpha}$ obey the usual anticommutation relations, $\{\tilde{a}_{p, \alpha}^{\dag}, \tilde{a}_{q, \beta}\} = \delta_{\alpha, \beta} \delta_{p, q}$.
Noting that the operator $P$ excludes the doubly occupied states, we do not have to consider the $F$-fermion Hilbert space where $F > 2d$ because doubly occupied states always appear in such a sector.
Therefore, in the following, we restrict ourselves to the subspaces with $2 \le F \le 2d$ and evaluate $h_{\mathrm{eff}}$.
Using the following relations 
\begin{align}
- P \tilde{a}_{p, \alpha}^{\dag} \tilde{a}_{q, \alpha} P 
&= \tilde{a}_{q, \alpha} P \tilde{a}_{p, \alpha}^{\dag} \ \ \text{for} \ p ,q \in \mathcal{E}_{o}, p\neq q, \\
P \left(1 - \sum_{\alpha=1}^{n} \tilde{a}_{p, \alpha}^{\dag} \tilde{a}_{p, \alpha} \right) P 
&= \tilde{a}_{p, \alpha} P \tilde{a}_{p, \alpha}^{\dag} \ \ \text{for} \ p \in \mathcal{E}_{o},
\end{align}
we have 
\begin{align}
h_{\mathrm{eff}}
=  \left(
E_{0} - 2s' nd + s'F (n-1)
\right) P 
+ s' \sum_{\alpha=1}^{n} \Psi_{\alpha} P \Psi_{\alpha}^{\dag}, \label{eq:heff}
\end{align}
where $\Psi_{\alpha}$ is defined as 
\begin{align}
\Psi_{\alpha} = \sum_{p \in \mathcal{E}_{o}} \tilde{a}_{p, \alpha}.
\end{align}
\par
We treat the two cases, $n \le d$ and $n > d$, separately.
We first discuss the former.
Let us denote the local ground state energy of $h_{\mathrm{eff}}$ in $\mathcal{H}_{F}(\Lambda_{o})$ by $E_{F}^{\mathrm{GS}}$.
Since the second term in Eq. (\ref{eq:heff}) is positive semidefinite, we see that 
\begin{align}
E_{F}^{\mathrm{GS}} \ge E_{0} - 2s' nd + s' F(n-1). \label{eq:lower bound 1}
\end{align}
Therefore, we obtain 
\begin{align}
\min_{n \le F \le 2d} E_{F}^{\mathrm{GS}} \ge E_{0} - 2s' nd + s' n(n-1). \label{eq:lower bound 2}
\end{align}
In fact, we can write down the local ground state of Eq. (\ref{eq:heff}) in $\mathcal{H}_{F}(\Lambda_{o})$ for $n \le F \le 2d$ using the method found in~\cite{brandt1992hubbard,verges1994ground} and find that $ E_{F}^{\mathrm{GS}} = E_{0} - 2s' nd + s' F(n-1)$ for $n \le F \le 2d$. 
See Appendix~\ref{app:2} for details.
For $F < n$, we note that in the direct sum decomposition (\ref{eq:J decomposition}), 
$\mathcal{J}_{M_{1}, \dots, M_{n}}$ has at least $(M-F)$ zero subscripts because $\sum_{\alpha=1}^{n} M_{\alpha} =F < n$.
Without loss of generality, we assume that $M_{F+1} = M_{F+2} = \cdots = M_{n} = 0$.
For any state $\ket{\psi} \in \mathcal{J}_{M_{1}, \dots, M_{F}, 0, \dots, 0}$, we see that 
\begin{align}
\sum_{\alpha=1}^{n} \sum_{p, q \in \mathcal{E}_{o}} P \tilde{a}_{p, \alpha}^{\dag} \tilde{a}_{q, \alpha} P \ket{\psi} 
&= \sum_{\alpha=1}^{F} \sum_{p, q \in \mathcal{E}_{o}} P \tilde{a}_{p, \alpha}^{\dag} \tilde{a}_{q, \alpha} P \ket{\psi},
\end{align}
and 
\begin{align}
P\left(1 - \sum_{\alpha=1}^{F} \tilde{a}_{p, \alpha}^{\dag} \tilde{a}_{p, \alpha}\right)P \ket{\psi} 
&= \tilde{a}_{p, \alpha} P \tilde{a}^{\dag}_{p, \alpha} \ket{\psi} \ \ \text{for} \ \alpha=1, \dots, F \ \text{and} \ p \in \mathcal{E}_{o}.
\end{align}
Using these relations, one finds that 
\begin{align}
h_{\mathrm{eff}} \ket{\psi}
=\left\{ \left( E_{0} - 2s' F d + s' F (F-1) \right) P + s' \sum_{\alpha=1}^{F} \Psi_{\alpha} P \Psi_{\alpha}^{\dag} \right\} \ket{\psi}. \label{eq:heff in J}
\end{align}
Therefore, the local ground state energy $E_{M_{1}, \dots, M_{F}, 0, \dots, 0}^{\mathrm{GS}}$ in $\mathcal{J}_{M_{1}, \dots, M_{F}, 0, \dots, 0}$ satisfies
\begin{align}
E_{M_{1}, \dots, M_{F}, 0, \dots, 0}^{\mathrm{GS}} \ge E_{0} - 2s' F d + s' F(F-1). \label{eq:lower bound 3}
\end{align}
Here, we note that the lower bound does not depend on the choice of $\mathcal{J}_{M_{1}, \dots, M_{n}}$.
From Eq. (\ref{eq:lower bound 3}), we obtain 
\begin{align}
E_{F}^{\mathrm{GS}} \ge E_{0} - 2s' Fd + s' F(F-1) \ \ \text{for} \ 2 \le F < n. \label{eq:lower bound 4}
\end{align}
Furthermore, the lower bound of Eq. (\ref{eq:lower bound 3}) is indeed saturated in $\mathcal{H}_{F}(\Lambda_{o})$ for $F < n$. 
See Appendix~\ref{app:2}.
Noting that $F < n \le d$, the right-hand side of Eq. (\ref{eq:lower bound 4}) takes the minimum value when $F = n-1$, and we find that 
\begin{align}
\min_{2 \le F < n} E_{F}^{\mathrm{GS}} \ge E_{0} - 2s' (n-1) d + s'(n-1)(n-2). \label{eq:lower bound 5}
\end{align}
Combining Eq. (\ref{eq:lower bound 2}) with Eq. (\ref{eq:lower bound 5}), we get 
\begin{align}
\min_{2 \le F \le 2d} E_{F}^{\mathrm{GS}} \ge E_{0} - 2s' n d + s' n (n-1), 
\end{align}
and its lower bound is non-negative when $0 < \nu < \nu_{\mathrm{c}}(n, d)$, where $\nu_{\mathrm{c}}(n, d)$ is defined as 
\begin{align}
\nu_{\mathrm{c}}(n,d) = \sqrt{\frac{2d+1 + \sqrt{4(2d-n)(n-1) + (2d+1)^{2}}}{2(2d-n)(n-1)}} \ \ \text{for} \ \ n \le d. 
\end{align}
\par
We next consider the case where $n > d$.
In this case, we further discuss the following two cases, $d < n \le 2d$ and $2d < n$.
When $d < n \le 2d$ and $n \le F \le 2d$, by repeating the previous argument, we obtain
\begin{align}
\min_{n \le F \le 2d} E_{F}^{\mathrm{GS}} \ge E_{0} - 2s' n d + s' n(n-1). \label{eq:lower bound 6}
\end{align}
On the other hand, for $d < n \le 2d$ and $2 \le F < n$, we have
\begin{align}
E_{F}^{\mathrm{GS}} \ge E_{0} - 2 s' F d + s' F(F-1) \ \ \text{for} \ 2 \le F < n .\label{eq:lower bound 7}
\end{align}
Since $n > d$, the lower bound of Eq. (\ref{eq:lower bound 7}) has the minimum value at $F=d$.
Therefore, we find that 
\begin{align}
\min_{2 \le F < n} E_{F}^{\mathrm{GS}} \ge E_{0} - s' d(d+1). \label{eq:lower bound 8}
\end{align}
Subtraction of the right-hand side of Eq. (\ref{eq:lower bound 8}) from the right-hand side of Eq. (\ref{eq:lower bound 6}) yields $s'(n-d)(n-d-1)$, which is non-negative because $n > d$.
Thus, we obtain the following inequality
\begin{align}
\min_{2 \le F \le 2d} E_{F}^{\mathrm{GS}} \ge E_{0} - s' d(d+1) \ \ \text{for} \ d < n \le 2d.
\end{align}
When $n >2d$, for all $F$ such that $2 \le F \le 2d$, it holds that $F < n$, and so we have 
\begin{align}
\min_{2 \le F \le 2d} E_{F}^{\mathrm{GS}} \ge E_{0} - s' d(d+1) \ \ \text{for} \ n> 2d.
\end{align}
Summarizing the above inequalities, we obtain  
\begin{align}
\min_{2 \le F \le 2d} E_{F}^{\mathrm{GS}} \ge E_{0} - s' d(d+1) \ \ \text{for} \ n > d, \label{eq:lower bound 9}
\end{align}
and find that the lower bound of Eq. (\ref{eq:lower bound 9}) is non-negative if $0 < \nu < \nu_{\mathrm{c}}(n, d)$, where 
\begin{align}
\nu_{\mathrm{c}}(n, d) = \sqrt{\frac{2d+1 + \sqrt{8d^{2} + 1}}{2d(d-1)}}.
\end{align}
Thus, we have proved Lemma~\ref{lemma:2}.
\hspace{\fill} $\square$
}
\par
Finally, Theorem~\ref{thm:2} can be proved using Lemma~\ref{lemma:1} and~\ref{lemma:2}.
We note that $h_{p}$ can be regarded as a finite-dimensional matrix independent of the system size since the local Hamiltonian acts nontrivially only on $4d+1$ sites.
This means that the energy levels of $h_{p}$ depend continuously on the parameters.
Therefore, Lemma~\ref{lemma:2} ensures that $h_{p}$ is positive semidefinite when $t/s$ and $U/s$ are finite but sufficiently large.
Lemma~\ref{lemma:1} implies that the ground states of $H_{2}$ are fully polarized states, which proves Theorem~\ref{thm:2}.
\hspace{\fill}$\square$

\begin{acknowledgements}
H. K. was supported in part by JSPS Grant-in-Aid for Scientific Research on Innovative Areas No. JP20H04630, JSPS KAKENHI Grant No. JP18K03445, and the Inamori Foundation.
\end{acknowledgements}

\begin{appendices}

\makeatletter  
\def\@seccntformat#1{\appendixname \,
\csname the#1\endcsname\sectcounterend: \,}
\makeatother

\section{The linear independence of many-body states \label{app:1}} 
Here, we show the linear independence of the states of the form 
\begin{align}
\ket{\Phi_{\{A_{\alpha}\}, \{B_{\alpha}\}}} = \left(\prod_{p \in A_{1}} a_{p,1}^{\dag}\right) \cdots \left(\prod_{p \in A_{n}} a_{p,n}^{\dag}\right) \left(\prod_{u \in B_{1}} b_{u, 1}^{\dag}\right) \cdots  \left(\prod_{u \in B_{n}} b_{u, n}^{\dag}\right) \ket{\Phi_{\mathrm{vac}}}, \label{eq:many-body a b state}
\end{align}
where $\{A_{\alpha}\}$ and $\{B_{\alpha}\}$ are arbitrary subsets of $\mathcal{E}$ and $\mathcal{I}$, respectively.
Let us denote by $\mathfrak{h} \cong \mathbb{C}^{|\Lambda|}$ the single-particle Hilbert space.
We define two types of localized states $\spstate{a}_{p} = (\spstate{a}_{p}(x))_{x \in \Lambda}$ and $\spstate{b}_{u} = (\spstate{b}_{u}(x))_{x \in \Lambda}$ in $\mathfrak{h}$ as 
\begin{align}
\spstate{a}_{p}(x) = \begin{cases}
1 & \ \ \text{if} \ x=p, \\
-\nu & \ \ \text{if} \ |x-p|=1/2, \\
0 & \ \ \text{otherwise},
\end{cases}
\end{align}
for each $p \in \mathcal{E}$, and 
\begin{align}
\spstate{b}_{u}(x) = \begin{cases}
1 & \ \ \text{if} \ x=u, \\
\nu & \ \ \text{if} \ |x-u|=1/2, \\
0 & \ \ \text{otherwise},
\end{cases}
\end{align}
for each $u \in \mathcal{I}$.
In terms of $\spstate{a}_{p}(x)$ and $\spstate{b}_{u}(x)$, the operators $a_{p, \alpha}$ and $b_{u, \alpha}$ can be written as 
\begin{align}
a_{p, \alpha} = \sum_{x \in \Lambda} \spstate{a}_{p}(x) c_{x, \alpha}, \\
b_{u, \alpha} = \sum_{x \in \Lambda} \spstate{b}_{u}(x) c_{x, \alpha}.
\end{align}
The states $\{\spstate{a}_{p}\}_{p \in \mathcal{E}}$ are linearly independent because for an arbitrary $p \in \mathcal{E}$, only $\spstate{a}_{p}$ has a nonzero component on site $p$.
Therefore, the Gram matrix $\mathsf{A}= \left(\mathsf{A}_{p, q}\right)_{p, q \in \mathcal{E}}$ defined by 
\begin{align}
\mathsf{A}_{p, q} = \sum_{x \in \Lambda} \spstate{a}_{p}(x) \spstate{a}_{q}(x)
\end{align}
is a regular matrix.
Now we introduce the dual operator of $a_{p, \alpha}$ as
\begin{align}
\bar{a}_{p, \alpha} = \sum_{q \in \mathcal{E}} \left(\mathsf{A}^{-1}\right)_{p, q} a_{q, \alpha},
\end{align}
for which we can check that 
\begin{align}
\{\bar{a}_{p, \alpha}^{\dag}, a_{q, \beta}\} = \delta_{\alpha, \beta} \delta_{p, q}. \label{eq:app dual a ac rel}
\end{align}
Similarly, since the states $\{\spstate{b}_{u}\}_{u \in \mathcal{I}}$ are also linearly independent, the Gram matrix $\mathsf{B} = \left(\mathsf{B}_{u, v}\right)_{u, v \in \mathcal{I}}$ defined by 
\begin{align}
\mathsf{B}_{u, v} = \sum_{x \in \Lambda} \spstate{b}_{u}(x) \spstate{b}_{v}(x)
\end{align}
is a regular matrix.
We define the dual operator of $b_{u, \alpha}$ as 
\begin{align}
\bar{b}_{u, \alpha} = \sum_{v \in \mathcal{I}} \left(\mathsf{B}^{-1}\right)_{u, v} b_{v, \alpha},
\end{align}
and see that 
\begin{align}
\{\bar{b}_{u, \alpha}^{\dag}, b_{v, \beta}\} = \delta_{\alpha, \beta} \delta_{u, v}. \label{eq:app dual b ac rel}
\end{align}
Having introduced the dual operators, we now prove the linear independence of the states Eq. (\ref{eq:many-body a b state}).
Suppose that
\begin{align}
\sum_{\{A_{\alpha}\}, \{B_{\alpha}\}} f\left(\{A_{\alpha}\}, \{B_{\alpha}\}\right) \ket{\Phi_{\{A_{\alpha}\}, \{B_{\alpha}\}}} = 0. \label{eq:app eq}
\end{align}
For arbitrary $\{A_{\alpha}\}$ and $\{B_{\alpha}\}$, we operate Eq. (\ref{eq:app eq}) from the left with \begin{align}
\left(\prod_{p \in A_{1}} \bar{a}_{p, 1}\right) \cdots \left(\prod_{p \in A_{n}} \bar{a}_{p, n}\right) \left(\prod_{u \in B_{1}} \bar{b}_{u, 1}\right) \cdots \left(\prod_{u \in B_{n}} \bar{b}_{u, n}\right), 
\end{align}
and then, using the anticommutation relations (\ref{eq:app dual a ac rel}) and (\ref{eq:app dual b ac rel}), we have $f\left(\{A_{\alpha}\}, \{B_{\alpha}\}\right) = 0$.
Thus, the states $\ket{\Phi_{\{A_{\alpha}\}, \{B_{\alpha}\}}}$ are linearly independent.
\section{The local ground state \label{app:2}} 
In this section, we construct the local ground state of Eq. (\ref{eq:heff}) in $\mathcal{H}_{F}(\Lambda_{o})$ explicitly.
To do this, we employ the method used in~\cite{brandt1992hubbard,verges1994ground}.
For $n \le F \le 2d$, we define a state $\ket{\Psi_{F}} \in \mathcal{H}_{F}(\Lambda_{o})$, as 
\begin{align}
\ket{\Psi_{F}} = \mathcal{N} P \left(\prod_{\alpha=1}^{n} \Psi_{\alpha}^{\dag} \right) \left(\prod_{q \in \{q_{1}, \dots, q_{F-n}\} \subset  \mathcal{E}_{o}} \tilde{a}_{q, \alpha_{q}}^{\dag} \right)\ket{\Phi_{\mathrm{vac}}}, 
\end{align}
where $q_{1}, \dots, q_{F-n}$ are distinct lattice sites in $\mathcal{E}_{o}$ and $\alpha_{q}$ can take an arbitrary color.
Here $\mathcal{N}$ denotes a normalization constant.
From the operator identity 
\begin{align}
P \Psi_{\alpha}^{\dag} P \Psi_{\alpha}^{\dag} = 0, \label{eq:operator identity}
\end{align} 
one finds that 
\begin{align}
s' \sum_{\alpha=1}^{n} \Psi_{\alpha} P \Psi_{\alpha}^{\dag} \ket{\Psi_{F}} = 0, 
\end{align}
which means that the state $\ket{\Psi_{F}}$ is the local ground state of Eq. (\ref{eq:heff}) in $\mathcal{H}_{F}(\Lambda_{o})$ when $F \ge n$.
We thus conclude that 
\begin{align}
E_{F}^{\mathrm{GS}} = E_{0} - 2 s' nd + s' F(n-1) \ \ \text{for} \ n \le F \le 2d.
\end{align}
\par
Next, we discuss the case when $2 \le F < n$.
In this case, it suffices to consider the effective Hamiltonian $h_{\mathrm{eff}}$ within the subspace $\mathcal{J}_{M_{1}, \dots, M_{F}, 0, \dots, 0}$, in which $h_{\mathrm{eff}}$ acts as Eq. (\ref{eq:heff in J}). 
As in the previous case, we can construct the local ground state in $\mathcal{J}_{M_{1}=1, \dots, M_{F}=1, 0, \dots, 0}$ as 
\begin{align}
\ket{\Psi} = \mathcal{N}' P \left(\prod_{\alpha=1}^{F} \Psi_{\alpha}^{\dag}\right) \ket{\Phi_{\mathrm{vac}}},
\end{align}
where $\mathcal{N}'$ is a normalization constant.
Again, from the operator identity (\ref{eq:operator identity}), we have $s' \sum_{\alpha=1}^{F} \Psi_{\alpha} P \Psi_{\alpha}^{\dag} \ket{\Psi} = 0$ and then obtain
\begin{align}
E_{F}^{\mathrm{GS}} = E_{0} - 2s' F d + s' F(F-1).
\end{align}
\end{appendices}

%
%

\bibliographystyle{spmpsci}      
\bibliography{reference}   

%
%


\end{document}